\documentclass[a4,12pt]{article}
\usepackage[dvips]{graphicx}
\usepackage{amsmath,amssymb}
\usepackage{amssymb}
\usepackage{enumerate}
\usepackage[usenames]{color}

\renewcommand{\thefootnote}{\fnsymbol{footnote}}

\begin{document}

\title{Interactions of Non-Abelian Global Strings}

\maketitle

\begin{center}
\author{
Eiji Nakano$^a$\footnote{e-mail: {\tt enakano(at)ntu.edu.tw}}, 
Muneto Nitta$^b$\footnote{e-mail: {\tt nitta(at)phys-h.keio.ac.jp}}
 and 
Taeko Matsuura$^c$\footnote{e-mail: {\tt taeko(at)ect.it}}
}

\bigskip\bigskip\bigskip
$^a$ {\it Department of Physics and Center for
Theoretical Sciences, National Taiwan University, Taipei 10617, Taiwan}\\

$^b$ {\it Department of Physics, Keio University, Hiyoshi, Yokohama,
Kanagawa 223-8521, Japan}\\

$^c$ {\it ECT*, Villa Tambosi, 
        strada delle Tabarelle, 286, I 38050 Villazzano (TN), Italy}\\

\abstract
{Non-Abelian global strings are expected to form
during the chiral phase transition. 
They have orientational zero modes in the internal space,
associated with the vector-like symmetry $SU(N)_{\rm L+R}$ 
broken in the presence of strings. 
The interaction among two parallel
non-Abelian global strings 
is derived for general relative orientational zero modes, 
giving a non-Abelian generalization of the Magnus force.
It is shown that
when the orientations of the strings are the same, 
the repulsive force reaches the maximum, 
whereas when the relative orientation becomes the maximum, 
no force exists between the strings.
For the Abelian case we find a finite volume correction 
to the known result. 
The marginal instability of the previously known 
Abelian $\eta'$ strings is discussed.
}

\end{center}

\newpage

\setcounter{footnote}{0}
\renewcommand{\thefootnote}{\arabic{footnote}}

\section{Introduction}


\indent

Topological strings play very important roles in physics.
Their study ranges from the cosmic strings which are formed 
in the early universe \cite{Vilenkin} to the vortices   
in the condensed matter physics $i.e.$ Abrikosov flux tubes 
in type II superconductors, 
the superfluid vortices in $^4$He and cold atoms, and so on. 
They are accompanied with the spontaneous symmetry breaking 
at phase transitions.  
In high energy physics,  
topological strings appear in the standard model, 
GUTs and other particle models \cite{Achucarro:1999it}.
In cosmology 
they had long been the strong candidates 
for the formation of the galaxies 
and CMB fluctuations. Although this possibility is now excluded, 
the study of cosmic strings is 
under the significant developments recently 
due to several reasons \cite{recent}.
Depending on whether a broken symmetry is global or local, 
strings are called global or local, respectively.
In the early stage of developments, 
global cosmic strings were not focused 
because their energy is logarithmically divergent. 
Later it was recognized that the divergence is not a problem 
because a finite volume system or nearest strings give 
a natural infrared cutoff. 
Then their interaction, reconnection (intercommutation) and 
formation of a network 
were extensively discussed 
\cite{Vilenkin:1982ks,Shellard:1987bv,Perivolaropoulos:1991du,Vilenkin}.


One of the important recent 
developments concerns the non-Abelian strings. 
Here we use the term ``non-Abelian string" for 
a string which arises at the symmetry breaking $G \to H$ for 
which the unbroken subgroup $H$ is non-Abelian. 
Recently the non-Abelian local/semi-local strings have been found 
in superstring theory \cite{Hanany:2003hp} and in supersymmetric QCD 
\cite{Auzzi:2003fs}. 
Since these strings are BPS 
{\it i.e.} at the critical coupling, 
no static force exists  
and the so-called moduli matrix approach 
\cite{Isozumi:2004vg,Eto:2006pg}
provides the most generic solutions 
and their complete moduli space \cite{Eto:2005yh}.
Their interaction, scattering and reconnection have been 
studied in the moduli space approximation \cite{Eto:2006db}. 
Non-Abelian semi-local strings have been further 
studied \cite{Eto:2006pg,Shifman:2006kd}. 


In contrast to these remarkable developments, 
the non-Abelian {\it global} strings 
have not been so much investigated yet, 
as was so in the case of the Abelian global strings.
Despite this, they are interesting for several reasons. 
First, different from the Abelian global strings, 
the non-Abelian strings have 
the internal degrees of freedom which are called $orientation$;
the presence of a string breaks the symmetry $H$ further 
$H \to H'$ and consequently the zero modes corresponding to $H/H'$
appear along the string. 
Then we have a continuously 
infinite number of strings with the same tension 
which are parameterized by 
this $orientation$, namely a point in $H/H'$.
The interaction among the strings with different orientations
is not trivial at all, 
which is the main issue of the present Letter. 
Second, it was shown that non-Abelian global strings 
with domain walls indeed form 
during the chiral phase transition 
in QCD \cite{Balachandran:2002je} 
where the $SU(N)_{\rm L} \times SU(N)_{\rm R} 
(\times {\bf Z}_{N,{\rm A}})$ symmetry 
is broken to its diagonal subgroup $SU(N)_{\rm L+R}$
where $N$ indicates the number of flavors.
In Ref.~\cite{Balachandran:2002je}, they explicitly took into account 
the effect of anomaly.

Before the discovery of the non-Abelian global strings in QCD, 
it was already shown that the Abelian global strings 
which are called the $\eta'$ strings may
exist in the early universe \cite{Brandenberger:1998ew}. 
When the temperature becomes very high, 
the chiral anomaly is not effective \cite{Pisarski:1983ms}. 
It is because the instantons require both color electric and 
magnetic fields. But the fluctuation of the electric field
is suppressed at high temperature due to the Debye screening.  
Therefore, if the temperature for the chiral 
phase transition is so high that the $U(1)_{\rm A}$ symmetry 
is effectively restored\footnote{
This point is still controversial and is not settled yet.
See Ref.~\cite{anomaly}, for example.}, 
the Abelian strings arise during the spontaneous 
symmetry breaking of this effective $U(1)_{\rm A}$ symmetry due to the chiral
condensate $\langle \bar{q} q \rangle $, where $q(\bar{q})$ indicates the (anti)quark fields.
The $\eta'$ strings become unstable as the temperature decreases and 
the instanton effects become substantial. 
The authors in Ref.~\cite{Balachandran:2001qn} expected that
they can be stable 
if they accompany three domain walls. 

In this Letter, we consider non-Abelian global strings which 
arise during the chiral phase transition when we can 
neglect the effect of the anomaly, which is just the 
case considered in Refs.~\cite{Brandenberger:1998ew, Balachandran:2001qn}. 
We derive the interactions among 
the non-Abelian global strings in the $U(N)_L \times U(N)_R$ linear 
$\sigma$ model. There are lots of interesting questions
about the formation, evolution of the strings etc. 
As a first step, however, we consider the interaction among the static two 
non-Abelian 
strings with various relative orientations using the Abrikosov 
approximation. In section 2, the non-Abelian string solution with 
general orientation is constructed in the $U(N)_L \times U(N)_R$ linear 
$\sigma$ model.  The interaction among the static two non-Abelian 
strings is derived in section 3. 
In the case when the orientations of two strings are the same 
the calculation reduces to that of two Abelian strings. 
We find even for this case a finite-volume correction 
to the known result 
\cite{Vilenkin:1982ks,Shellard:1987bv,Perivolaropoulos:1991du,Vilenkin}. 
We end in section 4 with conclusion 
and discussion.


\section{Non-Abelian global strings and orientations}

Let us consider the chiral $U(N)_{\rm L} \times U(N)_{\rm R}$ linear 
$\sigma$ model. We first introduce the $N$ by $N$ matrix field 
$\Phi_{ij}$ ($i,j=1 \cdots N$) in order to parameterize 
the symmetry breaking. This field belongs to $[N, \bar{N}]$
representation of $SU(N)_{\rm L} \times SU(N)_{\rm R}$.
Under the global chiral symmetry $G=SU(N)_{\rm L} \times SU(N)_{\rm R} 
\times U(1)_{\rm A}$, $\Phi$ transforms as 
\begin{eqnarray}
\Phi \to e^{i\alpha} U_L \Phi U_R^{\dag}.
\end{eqnarray}
where $U_L$ and $U_R$ are independent $SU(N)$ matrices
and $e^{i\alpha}$ is the total $U(1)_{\rm A}$ rotation.

The Lagrangian which is symmetric under $G$ is 
\begin{eqnarray}\label{GL}
{\cal L}=
 {\rm{tr}} ({\vec \partial}\Phi^{\dag} {\vec \partial}\Phi)
-m^2{\rm{tr}} (\Phi^{\dag}\Phi)
-\lambda_1 ( {\rm{tr}}\Phi^{\dag}\Phi)^2
-\lambda_2 {\rm{tr}} [(\Phi^{\dag}\Phi)^2] 
\end{eqnarray}
up to quartic order in $\Phi$. 
When $m^2 <0$, $\lambda_1 + \lambda_2 /N >0 $, and $\lambda_2 >0$, 
the vacuum expectation value 
\begin{eqnarray}
\langle \Phi \rangle = v {\bf 1} \equiv \Phi_0, 
~ v = \sqrt{-m^2/2(N \lambda_1 + \lambda_2)}
\end{eqnarray}
breaks the symmetry $G$ to $H=SU(N)_{\rm L+R} \times {\bf Z}_N$ 
and corresponding $N^2$ Nambu-Goldstone bosons appear. 
The action of $H$ to $\langle \Phi \rangle $ is 
$\langle \Phi \rangle \to e^{i\alpha} U_L 
\langle \Phi \rangle U_R^{\dag}$ with 
$(e^{i\alpha}, U_L, U_R )= (\omega ,\omega^{-1} U, U):
\omega \in {\bf Z}_N, U \in SU(N)$. 
The coset space has the non-trivial first homotopy group,  
\begin{equation} 
\frac{G}{H} = \frac{SU(N) \times U(1)}{{\bf Z}_N}=U(N) \ \quad 
\Rightarrow \ \quad 
\pi_1\left[U(N)\right]={\bf Z}, 
\end{equation}
which develops both the non-Abelian as well as Abelian strings. 
Therefore, the non-Abelian vortex strings 
we are studying here are topological objects 
contrary to the pion string which is non-topological 
with $\pi_1\left[SU(2)\right]=0$ \cite{Zhang:1997is}. 

We will consider the cylindrically symmetric string configuration 
along the $z$-axis.
The most fundamental string is the non-Abelian string which is generated by
both $SU(N)$ and $U(1)$ generators of $G$. 
At large distance from 
the core of the fundamental string, 
the matrix field $\Phi(\theta)$ rotates as:
\begin{eqnarray}\label{reference}
\Phi(\theta,r)&=& \exp \left(i \frac{\theta}{N}\right) 
\exp \left(-iT_{N^2-1} \frac{\sqrt{N(N-1)}}{N}\theta \right) \Phi(0, r)
\nonumber\\
&=& {\rm diag}\left( e^{i\theta} f(r), \, g(r),\cdots , \, g(r), \, g(r) \right) 
\nonumber \\
&\Rightarrow& {\rm diag}\left( e^{i\theta}, 1,\cdots , 1, 1 \right), 
\end{eqnarray}
where we have already taken $v=1$ for simplicity, 
and $T_a$ $(a=1,2, \cdots N^2-1)$ is the generators of $SU(N)$ in the 
fundamental representation which we normalize as
${\rm Tr}\{T_aT_b\}=\delta_{ab}$.
The ($N^2-1$)-th generator is 
$T_{N^2-1}= \frac{1}{\sqrt{N(N-1)}}{\rm diag}(1- N , \cdots , 1 ,1)$. 
Here $\theta$ is the angular coordinate in the $x$-$y$ plane 
and we set $\Phi(0,r)=\Phi_0$. 
The full numerical solution of the string with profile functions $f(r)$ and $g(r)$
is given in Ref.~\cite{NS}.

The string configuration breaks the symmetry $H$ further as 
$SU(N)_{\rm L+R} \to SU(N-1)_{\rm L+R} \times U(1)_{\rm L+R}$. 
Consequently the zero modes corresponding to 
\begin{eqnarray}
 {SU(N)_{\rm L+R} \over SU(N-1)_{\rm L+R} \times U(1)_{\rm L+R}} 
 \simeq {\bf C}P^{N-1}  \label{eq:orientation}
\end{eqnarray}
appear along the string.
Eq.~(\ref{reference}) is in fact just one 
particular string among a continuously
infinite number of strings with the same tension 
which are parameterized by 
the $orientation$, namely a point in ${\bf C}P^{N-1}$. 
We will explicitly construct the two 
string system with general relative orientation in the next section.

Before going to the next section 
we discuss stability of our solutions here.
Regarding dynamical stability of non-Abelian strings, 
the scaling argument from the Derrick theorem \cite{Derrick:1964ww} can not be applied 
to the present case, because they are global strings 
whose energy diverges in infinite size systems. 
They usually exist in finite size systems, 
as $U(1)$ vortex does in Helium superfluid, 
where the effect from boundary prevents vortex core from collapsing. 
In this meaning, 
once a cutoff parameter has been introduced for spatial boundary 
to make the total energy finite, 
a modified version of the Derrick theorem makes sense, 
see e.g., \cite{Perivolaropoulos:1992kf}. 
In contrast to texture-like 
objects discussed in \cite{Perivolaropoulos:1992kf} 
where only gradient energy terms are taken into consideration, 
our global string has a non-trivial stable solution 
where gradient and potential energies are balanced with a finite cutoff $\Lambda$. 
Also, there is an issue whether or not the global non-Abelian string solution
is stiff against small perturbations of diagonal elements
into which a vortex solution is not embedded. 
Here we would briefly show this stability: 
first introduce a small perturbation field $\psi(r,t)$ 
as $g(r) \rightarrow g(r)+\psi(r,t)$, 
and suppose that $\psi(r,t)=e^{-i\omega t} \psi(r)$. 
Plugging this into the equation of motion 
and linearizing it in $\psi(r)$ lead to a 
Schr\"{o}dinger-like equation. 
After a normalization, we obtain 
\begin{eqnarray}
\psi''+\frac{1}{r}\psi'
+\left[ 1+\omega^2-2\kappa f^2(r)-2\left( \kappa (N-1)+1\right) 3 g^2(r) \right] \psi=0. 
\end{eqnarray}
We solve an eigenvalue problem for $\omega^2$ 
with the boundary condition $\psi'(0)=\psi'(\Lambda)=0$. 
If all the eigenvalues of $\omega^2$ are 
positive for given $\kappa$ and $N$, 
the string solution is stable. 
$1)$ In the case of $\kappa=0$ (the critical coupling), 
$f(r)$ and $g(r)$ are decoupled and 
it is immediately found that 
$g(r)=1/2$, and then 
$\psi''+\frac{1}{r}\psi'
+\left[ 1+\omega^2-6 g(r)^2 \right] \psi=0$. 
The Bessel function gives the solution 
and only positive $\omega^2$'s satisfy 
the boundary condition. 
$2)$ The case of $\kappa\neq 0$ is more complicated. 
After the substitution of the full solutions 
for $f(r)$ and $g(r)$ numerically obtained in 
Ref.~\cite{NS}, 
for instance 
for $\kappa=0.2$ and $N=3$, 
we found $\omega^2=1.9851$ as the lowest eigen value.
We thus see that the non-Abelian string solutions 
are dynamically stable as expected from the topology arguments.
The complete analysis on the stability is beyond 
the scope of the present paper.

In the situation that the $U(1)_A$ symmetry in our Lagrangian is gauged, 
our non-Abelian strings become semi-local strings 
(with finite energy) \cite{Vachaspati:1991dz}, 
then dynamical stability mechanism by Hindmarsh \cite{Hindmarsh:1991jq}, 
which is related to magnitudes between gauge and scalar couplings, 
might work even in infinite size systems. 
Although the vortex solution discussed 
in \cite{Hindmarsh:1991jq} is not topological, 
Hindmarsh has also mentioned the instability arising 
from small fluctuation of the solution. 
But this is not our present case with global $U(1)_{\rm A}$.

\section{Interaction between two strings}

\begin{figure}
\begin{center}
\includegraphics[height=3.0cm, width=6cm]{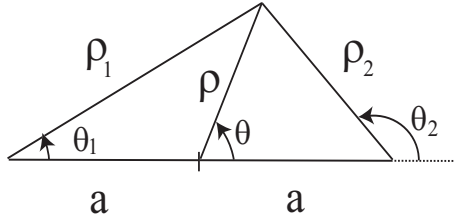}
\end{center}
\caption{\label{fig1} 
Configuration of two global strings with interval $d=2a$ 
in polar coordinate $(\rho, \theta)$.}
\end{figure}

Now we consider the interaction among arbitrary two strings. 
Let us place two strings $\phi_{1,2}$ 
parallel along the $z$-axis 
with the separation $2a$ in the $x$-$y$ plane.
For definiteness those positions are 
$(\rho, \theta)=(a,0)$ and $(a,\pi)$ 
as in Fig.~\ref{fig1} 
where  
$(\rho, \theta)$ are the polar coordinates in the $x$-$y$ plane.
As the orientation in the internal space 
${\bf C}P^{N-1}$ in (\ref{eq:orientation}) is concerned, 
only the relative orientation matters.
Let us take the reference string $\phi_1$ as in Eq.~(\ref{reference}):
\begin{eqnarray}\label{phi1}
 \phi_1  = {\rm diag}(e^{i\theta_1},1, \cdots , 1, 1).
 \label{phi1}
\end{eqnarray} 
Then starting from the same orientation with $\phi_1$ in 
(\ref{phi1}), 
the most general orientation (\ref{eq:orientation}) for 
the second string $\phi_2$ 
is obtained by acting $SU(N)_{\rm L+R}$ on it.
However as far as two string interaction is concerned, 
only an $SU(2)_{\rm L+R} (\subset SU(N)_{\rm L+R})$ rotation 
is enough to be considered without loss of generality.
(This corresponds to considering 
a ${\bf C}P^1$ submanifold inside the whole ${\bf C}P^{N-1}$.) 
We thus have 
\begin{eqnarray}
\phi_2 = \left(\begin{array}{cc}
g
\left(\begin{array}{cc}
e^{i\theta_2} & 0 \\
0 & 1 \\  
\end{array} \right)
g^{-1}
& 0 \nonumber \\
0 & {\bf 1}_{N-2} \nonumber \\
 \end{array} \right),
\end{eqnarray}
where $g$ is an element of $SU(2)$:
\begin{eqnarray}
g = \cos \left({\alpha\over 2} \right) {\bf 1}_2 
   + i \vec n \cdot \vec{\sigma} \sin \left({\alpha \over 2}\right) 
\end{eqnarray}
with $\vec \sigma=(\sigma_1,\sigma_2,\sigma_3)$ the Pauli matrices 
and $\vec n$ a unit three vector.
Since the rotation by $\sigma_3$ does not change the relative
orientation between $\phi_1$ and $\phi_2$, $n_z$ is fixed to $0$.
If we define $\beta$ by 
\begin{eqnarray}
 e^{i \beta} = n_x - i n_y, 
\end{eqnarray}
then $\alpha$ and $\beta$ parameterize 
${\bf C}P^1 \simeq S^2$.
Consequently, $\phi_2$ is simplified as 
\begin{equation}\label{phi2}
\phi_2 = \left(\begin{array}{cc}
\left(\begin{array}{cc}
e^{i\theta_2} \cos ^2 \left({\alpha\over 2}\right) 
 +  \sin ^2 \left({\alpha\over 2}\right)
& \frac{i}{2}(1-e^{i\theta_2}) e^{i \beta} \sin \alpha  \\
-\frac{i}{2}(1-e^{i\theta_2}) e^{- i \beta} \sin \alpha 
& \cos ^2 \left({\alpha\over 2}\right) 
 + e^{i\theta_2} \sin ^2 \left({\alpha\over 2}\right)  \\  
\end{array} \right)
& 0  \\
0 & {\bf 1}_{N-2}  \\
 \end{array} \right).
\end{equation}
For $\alpha =0$ the orientations of the two strings 
become the same, and the problem is reduced to the one of 
the Abelian strings \cite{Shellard:1987bv,Perivolaropoulos:1991du}. 
$\phi_{1,2}$ becomes an anti-string by changing the sign of 
$\theta_{1,2}$.

Let us now calculate the interaction among two 
parallel non-Abelian (anti-)strings with general orientations 
in the internal space.
The interaction energy density of the two string system 
is obtained by subtracting two individual string energies 
from the total configuration energy:
\begin{eqnarray}\label{F1}
F(\rho,\theta,a,\alpha)=
{\rm tr}
\left( |\partial \Phi_{\rm tot}|^2 -|\partial \phi_1|^2
-|\partial \phi_2|^2  \right),
\end{eqnarray}
where $\Phi_{\rm tot}$ is the total string configuration and 
we have used the fact that for sufficiently large value of $a$ 
the potential energies can be approximated by 
$V(\Phi_{\rm tot})=V(\phi_1) = V(\phi_2) = 0$.
We employ the Abrikosov ansatz for the configuration where 
\begin{eqnarray}
 \Phi_{\rm tot} = \phi_1\phi_2 \;\;\;\;  
 {\rm or} \;\;\;\; \phi_2 \phi_1.
\end{eqnarray}
We see that either ansatz gives the same result, 
so we do not have to worry about the ordering of 
the matrices.\footnote{
One can show that 
${\rm tr} \partial (\phi_1 \phi_2) \partial (\phi_1 \phi_2) = 
{\rm tr} \partial (\phi_2 \phi_1) \partial (\phi_2 \phi_1) $ 
up to reparameterization of $g$ and coordinates. }
Further, for simplicity, $\phi_{1,2}$ and $\Phi_{\rm tot}$ 
are approximated to their values at spatial infinity, 
Eqs.~(\ref{phi1}, \ref{phi2}). This approximation 
is justified when the interval of the strings is 
much longer than the coherence length 
(the transverse size of strings \cite{NS}):
\begin{eqnarray}
 a \gg m^{-1}.
\end{eqnarray}
Then Eq.~(\ref{F1}) is simplified and we get:
\begin{eqnarray}
F(\rho,\theta,a,\alpha)= \pm 
 (1 + \cos \alpha) \left( 
\frac{-a^2 + \rho^2}
{a^4 + \rho^4 -2 a^2 \rho^2 \cos (2 \theta)}
\right).
\end{eqnarray}
Here and below, the upper(lower) sign corresponds to 
the interaction energy density
for the string-string (string-anti-string) configuration.
For $\alpha =0$, $F$ reduces to that of Abelian global strings 
\cite{Shellard:1987bv,Perivolaropoulos:1991du}. 
However in contrast to the results in 
Refs.~\cite{Shellard:1987bv,Perivolaropoulos:1991du}, 
we have got the $\theta$ dependent interaction energy density 
which reaches the maximum (minimum) at $\theta=0, \pi$
when $\rho > a \, (\rho < a)$ for string-string configuration.
The $\theta$ dependence gives a correction 
to \cite{Shellard:1987bv,Perivolaropoulos:1991du} 
for the Abelian case ($\alpha=0$). 

The (sum of) tension, the energy of the strings per unit length,  
is obtained by integrating the energy density
over the $x$-$y$ plane:
\begin{eqnarray}
E(a,\alpha,L)&=& \pm 
\int _{0}^{L} d\rho 
\int _{0}^{2 \pi} d\theta  \rho F(\rho,\theta,a) \nonumber \\
&=& \pm \pi (1 + \cos \alpha)
\left[ -\ln 4 -2 \ln a  + \ln \left( a^2 + L^2 \right)  \right],
 \label{eq:int-energy}
\end{eqnarray}
where the IR cutoff $L$ is introduced to make the integral finite. 
The force between the two (anti-)strings are then obtained by 
differentiating $E$ by the interval:
\begin{eqnarray}
 f(a,\alpha,L)= \mp  \frac{\partial E}{2 \partial a}
 = \pm (1 + \cos \alpha) 
\left(\frac{\pi}{a}
- \frac{\pi a}{a^2 + L^2}  \right) \simeq 
\pm (1 + \cos \alpha) \frac{ \pi }{a},
\end{eqnarray}
where the last expression is for $a \ll L \to \infty$.
This is just the force between two Abelian (anti-)strings 
known as the Magnus force, multiplied by 
$(1 + \cos \alpha)/2$.
\begin{figure}
\begin{center}
\includegraphics[width=7cm]{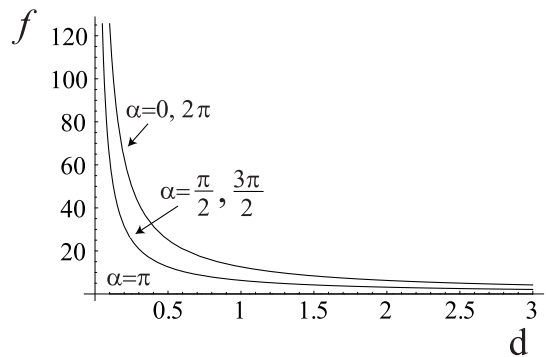}
\end{center}
\caption{\label{fig2} Dependence of the force between two non-Abelian strings
on the separation $d=2a$ for several $\alpha$. }
\end{figure}
We can see that when the orientation of the strings are the same 
$(\alpha=0)$ the repulsive(attractive) force reaches the maximum and 
is the same as that between Abelian global string and (anti-) string 
\cite{Shellard:1987bv,Perivolaropoulos:1991du}, 
where the second term in the middle equation 
gives a finite-volume (finite $L$) correction to 
\cite{Shellard:1987bv,Perivolaropoulos:1991du}. 
On the other hand, when the relative orientation
becomes the maximum ($\alpha=\pi$), no force exists between the strings.
Note that although the most stable configuration 
is given by 
the strings with the maximum relative angle ($\alpha=\pi$), 
it is not possible for the strings 
to change $\alpha$ because it is non-normalizable and 
must be fixed by the boundary condition at infinity.
This change is possible only if strings emit infinite number of 
Nambu-Goldstone bosons $\alpha$ in ${\bf C}P^{N-1}$.

So far
we have considered the case of the strings in infinite region 
where the relative orientation $\alpha$ is non-normalizable and is fixed. 
However 
$\alpha$ becomes a normalizable mode 
in a finite volume (finite $L$) which is realistic
in experiments such as the heavy-ion collider. 
In such a case, the force among orientations of 
two strings can be considered.
The interaction energy (\ref{eq:int-energy})
shows a repulsive force exists between 
aligned orientations of two strings. 
The stable configuration is for $\alpha = \pi$ 
where two orientations are anti-aligned. 
Therefore we conclude that they behave like antiferromagnet.

\section{Discussion}

\indent

In this Letter, we have considered the interactions among 
two non-Abelian strings in $U(N)_L \times U(N)_R$ linear $\sigma$ model.
This model also has an Abelian string solution 
\cite{Brandenberger:1998ew,Balachandran:2001qn}, the $\eta'$ string.
However, it is not the fundamental string and is made of 
$N$ non-Abelian strings:
\begin{eqnarray}
\Phi(\theta)&=& {\rm diag}(e^{i\theta}, \cdots , e^{i\theta}, e^{i\theta})
 \nonumber\\
&\sim&{\rm diag}(e^{i\theta}, \cdots , 1, 1)
\times {\rm diag}(1, e^{i\theta}, \cdots , 1)
\times \cdots 
\times {\rm diag}(1, 1,  \cdots , e^{i\theta}). 
\end{eqnarray}
There are no force among any of these non-Abelian strings, 
which indicates that the $\eta'$ string is marginally 
unstable to decay into $N$ non-Abelian strings. 
No binding energy implies that 
they decay with arbitrary momentum or by fluctuations.
This result holds 
in the presence of the chiral anomaly 
at lower temperatures; 
the Abelian string with
$N$ domain walls will decay into $N$ non-Abelian strings, 
where each is attached by one domain wall. 
In that case, the instability 
increases since once the Abelian string decays, 
the domain wall pulls the string away to infinity.
Therefore we do not have a cosmological domain wall problem.

The same type of the non-Abelian strings also appear  
in the low energy theory of supersymmetric QCD \cite{Auzzi:2003fs} and 
in the high density QCD (color superconductors) \cite{Balachandran:2005ev} 
as fundamental strings.
In these cases, the strings accompany the gauge fields 
which may change the interaction among them.
The case of strings in color superconductors is reported \cite{NNM} 
in which the universal repulsion is found 
unlike the case of global strings in this Letter.

Another interesting issue is how the non-Abelian strings 
emit or interact with the Nambu-Goldstone bosons 
(the pions and the $\eta'$ meson). 
In the case of global $U(1)$ strings, 
this can be described by using the two index antisymmetric 
tensor fields of the Kalb-Ramond action \cite{Kalb:1974yc}.
The non-Abelian tensor fields \cite{NA-2form}
may be suitable to describe the non-Abelian case.

Thermal effect was studied for non-Abelian local and semi-local 
vortices \cite{Eto:2007aw}. 
Finite temperature effect is important to study 
strings at a collider or in the early universe.

The inclusion of the bare quark mass would be a 
next step. If the quark mass enters in the theory, 
the chiral symmetry becomes not intact. 
Then a new topological object would appear 
where strings with different orientations 
are separated by bead-like solitons. 
Also, the ring-shaped string may appear.
We remain the study of these new topological 
objects as a future work.

\section*{Acknowledgements}
T.M. thanks S. Digal for helpful comments 
and encouragements. 
The authors are deeply grateful to Noriko Shiiki 
for providing us numerical data of the full string solutions. 
The work of E.N. is supported by 
Center for Theoretical Sciences, National Taiwan University 
under grant NO.~NSC96-2811-M-002-024.



\begin{thebibliography}{99}

\bibitem{Vilenkin}
A. Vilenkin and E. P. S. Shellard, Cosmic Strings
and Other Topological Defects, Cambridge Univ. Press
(1994); M. B. Hindmarsh and T. W. B. Kibble, Rept. \ 
Prog. \  Phys. \ {\bf 58}, 477 (1995).

\bibitem{Achucarro:1999it}
  A.~Achucarro and T.~Vachaspati,
  Phys.\ Rept.\  {\bf 327}, 347 (2000)
  [arXiv:hep-ph/9904229];
  R.~Jeannerot, J.~Rocher and M.~Sakellariadou,
  Phys.\ Rev.\ D {\bf 68}, 103514 (2003)
  [arXiv:hep-ph/0308134].


\bibitem{recent}
J.~Polchinski,
  arXiv:hep-th/0412244;
%
  T.~W.~B.~Kibble,
  arXiv:astro-ph/0410073.

\bibitem{Vilenkin:1982ks}
  A.~Vilenkin and A.~E.~Everett,
  Phys.\ Rev.\ Lett.\  {\bf 48}, 1867 (1982).

\bibitem{Shellard:1987bv}
  E.~P.~S.~Shellard,
  Nucl.\ Phys.\  B {\bf 283}, 624 (1987).

\bibitem{Perivolaropoulos:1991du}
  L.~Perivolaropoulos,
  Nucl.\ Phys.\  B {\bf 375}, 665 (1992).


\bibitem{Hanany:2003hp}
  A.~Hanany and D.~Tong,
  JHEP {\bf 0307}, 037 (2003)
  [arXiv:hep-th/0306150].

\bibitem{Auzzi:2003fs}
  R.~Auzzi, S.~Bolognesi, J.~Evslin, K.~Konishi and A.~Yung,
  Nucl.\ Phys.\  B {\bf 673}, 187 (2003)
  [arXiv:hep-th/0307287].


\bibitem{Isozumi:2004vg}
  Y.~Isozumi, M.~Nitta, K.~Ohashi and N.~Sakai,
  Phys.\ Rev.\  D {\bf 71}, 065018 (2005)
  [arXiv:hep-th/0405129];
  Phys.\ Rev.\ Lett.\  {\bf 93}, 161601 (2004)
  [arXiv:hep-th/0404198];
  Phys.\ Rev.\  D {\bf 70}, 125014 (2004)
  [arXiv:hep-th/0405194].

\bibitem{Eto:2006pg}
  M.~Eto, Y.~Isozumi, M.~Nitta, K.~Ohashi and N.~Sakai,
  J.\ Phys.\ A  {\bf 39}, R315 (2006)
  [arXiv:hep-th/0602170].

\bibitem{Eto:2005yh}
  M.~Eto, Y.~Isozumi, M.~Nitta, K.~Ohashi and N.~Sakai,
  Phys.\ Rev.\ Lett.\  {\bf 96}, 161601 (2006)
  [arXiv:hep-th/0511088];
  M.~Eto, K.~Konishi, G.~Marmorini, M.~Nitta, K.~Ohashi, W.~Vinci and N.~Yokoi,
  Phys.\ Rev.\  D {\bf 74}, 065021 (2006)
  [arXiv:hep-th/0607070].


\bibitem{Eto:2006db}
  M.~Eto, K.~Hashimoto, G.~Marmorini, M.~Nitta, K.~Ohashi and W.~Vinci,
  Phys.\ Rev.\ Lett.\  {\bf 98}, 091602 (2007)
  [arXiv:hep-th/0609214].


\bibitem{Shifman:2006kd}
  M.~Shifman and A.~Yung,
  Phys.\ Rev.\  D {\bf 73}, 125012 (2006)
  [arXiv:hep-th/0603134],
 M.~Eto {\it et al.},
 Phys.\ Rev.\  D {\bf 76}, 105002 (2007)
 [arXiv:0704.2218 [hep-th]].

\bibitem{Balachandran:2002je}
  A.~P.~Balachandran and S.~Digal,
  Phys.\ Rev.\  D {\bf 66}, 034018 (2002)
  [arXiv:hep-ph/0204262].

\bibitem{Brandenberger:1998ew}
  R.~H.~Brandenberger and X.~m.~Zhang,
  Phys.\ Rev.\  D {\bf 59}, 081301 (1999)
  [arXiv:hep-ph/9808306].

\bibitem{Pisarski:1983ms}
  R.~D.~Pisarski and F.~Wilczek,
  Phys.\ Rev.\  D {\bf 29}, 338 (1984).

\bibitem{anomaly}
  E.~V.~Shuryak,
  Comments Nucl.\ Part.\ Phys.\  {\bf 21}, 235 (1994);
  T.~D.~Cohen,
  Phys.\ Rev.\  D {\bf 54}, 1867 (1996);
  C.~W.~Bernard {\it et al.},
  Phys.\ Rev.\ Lett.\  {\bf 78}, 598 (1997);
  S.~Chandrasekharan, D.~Chen, N.~H.~Christ, W.~J.~Lee, R.~Mawhinney and P.~M.~Vranas,
  Phys.\ Rev.\ Lett.\  {\bf 82}, 2463 (1999);
  R.~A.~Janik, M.~A.~Nowak, G.~Papp and I.~Zahed,
  AIP Conf.\ Proc.\  {\bf 494}, 408 (1999);
  K.~Fukushima, K.~Ohnishi and K.~Ohta,
  Phys.\ Rev.\  C {\bf 63}, 045203 (2001);
  H.~Nagahiro, M.~Takizawa and S.~Hirenzaki,
  Phys.\ Rev.\  C {\bf 74}, 045203 (2006);
  D.~Horvatic, D.~Klabucar and A.~E.~Radzhabov,
  arXiv:0708.1260 [hep-ph].

\bibitem{Balachandran:2001qn}
  A.~P.~Balachandran and S.~Digal,
  Int.\ J.\ Mod.\ Phys.\  A {\bf 17}, 1149 (2002)
  [arXiv:hep-ph/0108086].

\bibitem{Zhang:1997is}
  X.~Zhang, T.~Huang and R.~H.~Brandenberger,
  Phys.\ Rev.\  D {\bf 58}, 027702 (1998)
  [arXiv:hep-ph/9711452].

\bibitem{Derrick:1964ww}
  G.~H.~Derrick,
  J.\ Math.\ Phys.\  {\bf 5}, 1252 (1964).

\bibitem{Perivolaropoulos:1992kf}
  L.~Perivolaropoulos,
  Phys.\ Rev.\  D {\bf 46}, 1858 (1992)
  [arXiv:hep-ph/9207256].

\bibitem{Vachaspati:1991dz}
 T.~Vachaspati and A.~Achucarro,
 Phys.\ Rev.\  D {\bf 44}, 3067 (1991).
 

\bibitem{Hindmarsh:1991jq}
  M.~Hindmarsh,
  Phys.\ Rev.\ Lett.\  {\bf 68}, 1263 (1992).


\bibitem{NS}
  M.~Nitta and N.~Shiiki,
  Phys.\ Lett.\  B {\bf 658}, 143 (2008)
  [arXiv:0708.4091 [hep-ph]].

\bibitem{Balachandran:2005ev}
  A.~P.~Balachandran, S.~Digal and T.~Matsuura,
  Phys.\ Rev.\  D {\bf 73}, 074009 (2006)
  [arXiv:hep-ph/0509276].


\bibitem{NNM}
  E.~Nakano, M.~Nitta and T.~Matsuura,
  Phys.\ Rev.\  D {\bf 78}, 045002 (2008)
  [arXiv:0708.4096 [hep-ph]].

\bibitem{Kalb:1974yc}
  M.~Kalb and P.~Ramond,
  Phys.\ Rev.\  D {\bf 9}, 2273 (1974).

\bibitem{NA-2form}
  K.~Seo, M.~Okawa and A.~Sugamoto,
  Phys.\ Rev.\  D {\bf 19}, 3744 (1979); 
  D.~Z.~Freedman and P.~K.~Townsend,
  Nucl.\ Phys.\  B {\bf 177}, 282 (1981).


\bibitem{Eto:2007aw}
 M.~Eto, T.~Fujimori, M.~Nitta, K.~Ohashi, K.~Ohta and N.~Sakai,
 Nucl.\ Phys.\  B {\bf 788}, 120 (2008)
 [arXiv:hep-th/0703197].

\end{thebibliography}
\end{document}